\documentclass[%
 reprint,longbibliography,preprintnumbers,
nofootinbib,
 amsmath,amssymb,
 aps,
pre,
]{revtex4-2}
\pdfoutput=1
\usepackage[utf8]{inputenc}
\usepackage{flushend}
\usepackage{bm}
\usepackage{balance}
\usepackage{tensor}
\usepackage{physics}
\usepackage{wrapfig}
\usepackage{subcaption}
\usepackage{graphicx}
\usepackage{dcolumn}
\usepackage{bm}

\usepackage[normalem]{ulem}

\usepackage[colorlinks = true,
            linkcolor = blue,
            urlcolor  = blue,
            citecolor =green,
            anchorcolor = blue]{hyperref}
\usepackage{verbatim}
\usepackage{color,ulem}
\usepackage[english]{babel}

\usepackage[utf8]{inputenc}
\input Starburst.fd
\newcommand*\initfamily{\usefont{U}{Starburst}{xl}{n}}\initfamily 

\newcommand{\beq}{\begin{eqnarray}}
\newcommand{\eeq}{\end{eqnarray}}
\usepackage{amsmath}
\usepackage{tikz}
\usetikzlibrary{decorations.pathmorphing}
\usetikzlibrary{shapes.misc}
\tikzset{cross/.style={cross out, draw=black, minimum size=8*(#1-\pgflinewidth), inner sep=0pt, outer sep=0pt},
cross/.default={1pt}}
\usetikzlibrary{patterns,math}
\begin{document}

\title{Relativistic theory of the viscosity of fluids across the entire energy spectrum}

\author{\textbf{Alessio Zaccone$^{1,2}$}}%
 \email{alessio.zaccone@unimi.it}

 \vspace{1cm}
 
\affiliation{$^{1}$Department of Physics ``A. Pontremoli'', University of Milan, via Celoria 16,
20133 Milan, Italy}

\affiliation{$^{2}$Institute of Theoretical Physics, University of G\"ottingen, Friedrich-Hund-Platz 1, 37077 G\"ottingen, Germany.}

\begin{abstract}
The shear viscosity is a fundamental transport property of matter. Here we derive a general theory of the viscosity of gases based on the relativistic Langevin equation (deduced from a relativistic Lagrangian) and nonaffine linear response theory. The proposed relativistic theory is able to recover the viscosity of non-relativistic classical gases, with all its key dependencies on mass, temperature, particle diameter and Boltzmann constant, in the limit of Lorentz factor $\gamma=1$. It also unveils the relativistic enhancement mechanism of viscosity. In the limit of ultrarelativistic fluids, the theory provides a new analytical formula which reproduces the cubic increase of viscosity with temperature in agreement with various estimates for hot dense matter and the QGP-type fluid.
\end{abstract}

\maketitle
\section{Introduction}
The viscosity is the key transport property of fluids, which determines their mechanical and thermal behaviour \cite{Murillo,Zaccone_viscosity}. In modern physics applications, viscosity is a key quantity to describe exotic states of matter such as quark-gluon plasmas \cite{quark,10.21468/SciPostPhys.10.5.118}, the microscopic mechanism of superfluidity in helium-4 \cite{Trachenko_2023}, and the dynamics of nuclear fission \cite{Sierk} and fusion-fission reactions \cite{Eccles,Vardaci,fusion-fission,Amano}. 
In cosmology and high-energy collision physics, the viscosity is a key coefficient that is required as input to compute the hydrodynamics of relativistic fluids \cite{Noronha,Ratti,Bemfica,Denicol_Noronha, Kovtun,Gavassino_2023,geroch2001,Hippert}. An even more fundamental problem, is that of the viscosity of a gas of relativistic hard-sphere particles \cite{Synge,Denicol_Noronha}. A number of approaches have been presented, which are all based on the relativistic Boltzmann equation solved either numerically \cite{Gabbana} or analytically for specific models \cite{Denicol}, typically within the Chapman-Enskog approximation. 


In this paper, we present a new simple analytical theory for the viscosity of classical particle systems valid throughout the energy spectrum, i.e. able to recover both the high-energy ultrarelativistic limit (with the correct temperature trend) as well as the well-known classical gas viscosity formula upon taking the limit $\gamma \rightarrow 1$. The approach is derived from first principles from a relativistic Lagrangian, which makes it useful for applications in high-energy physics, and provides new insights into the relativistic enhancement of viscosity via proper-momentum dissipation.

\section{Summary of previous work}
To summarize a rich landscape of results, the viscosity of gases can be parameterized in the following form \cite{Gabbana}:
\begin{equation}
    \eta = a P \tau.
\end{equation}
Here, $a$ is a prefactor of order unity; it is $a=4/5$ in the ultrarelativistic limit for an ideal gas in the relaxation time approximation (Anderson-Witting) \cite{Anderson,Rocha} and $a \rightarrow 1$ in the non-relativistic limit \cite{Ambrus}. Furthermore, $P$ is the pressure and $\tau$ is the characteristic relaxation time (e.g. the mean free time). 
In the non-relativistic limit, for an ideal gas $P \sim T$, the mean free time is given by $\tau=\ell/v$, with $\ell$ the mean free path and $v$ the average speed. Since $\ell$ is independent of $T$ and $v \sim \sqrt{T}$, one readily recovers the well-known result from the kinetic theory of gases \cite{Vincenti,Hildebrand}: $\eta \sim T^{1/2}$.

Turning to the high-energy fluids, the picture is much more fragmented. Depending on the cross-section assumptions used in a given theory, the mean-free time $\tau$ can take very different trends with $T$ \cite{Christiansen,Denicol}. Furthermore, in realistic systems, the cross section depends strongly on the momentum, which, in turn, depends on the fermionic or bosonic character of the mediator.
The same is true for the pressure $P$, the form of which, including its dependence on $T$, can be very different depending on the details of the theory (e.g. on whether pair-production processes are taken into account or not) and on the specific equation of state chosen. In spite of these differences for $\tau$ and $P$, most theories agree on the overall temperature dependence of the shear viscosity for high-energy fluids including the QGP, to be cubic $\eta \sim T^3$ \cite{Denicol_2012,Denicol,Christiansen}.

\section{Theory}
\subsection{Relativistic Caldeira-Leggett Lagrangian}
In what follows, we present a new theory, derived from a microscopic Lagrangian leading to a relativistic Langevin equation for the particle motion. Using linear response theory, this leads to a relativistic microscopic formula for the viscosity of real gases. Due to the explicit presence of the Lorentz $\gamma$ factor in this formula, the non-relativistic limit is correctly recovered for $\gamma \rightarrow 1$. In the ultrarelativistic limit of a high energy fluid, the theory correctly reproduces the $\sim T^{3}$ dependence in agreement with evidence and yields a new formula involving fundamental constants, mass and temperature.

The system is composed of a massive tagged particle (TP) dynamically coupled to a heat bath (HB) of relativistic oscillators (RO), each with mass $m_i$ and frequency $\omega_i$, which effectively represent the other degrees of freedom in the system in terms of the system's normal modes (the latter can of course be defined also for gases, see e.g. Ref. \cite{Kardar_2007}). 
The whole process is adiabatic, in the sense that the tagged particle only exchanges work with the heat bath by being dynamically coupled to each RO with a coupling constant $g_i$.
As the time parameter, we choose the lab time coordinate $t$, and denote with a dot the lab time derivative \cite{Petrosyan_2022,Zadra}.

The relativistic Caldeira-Leggett Lagrangian thus has two components for the tagged particle and the heat bath, respectively $L=L_{TP}+L_{HB}$ \cite{Petrosyan_2022,Zadra},
with
\begin{align}
		L_{\mathrm{TP}}&=\frac{mc^2}{\gamma(\dot{\vec{x}})}-V_{\mathrm{ext}}(\vec{x},\dot{\vec{x}},t),\\
		L_{\mathrm{HB}}&=\sum_i\frac{m_ic^2}{\gamma(\dot{\vec{q_i}})}-\frac{m_i \omega_i^2}{2\gamma(\dot{\vec{x}})}\norm{\vec{q}_i-\frac{g_i}{\omega_i^2}\vec{x}}_4^2 \label{interaction}
\end{align}
where we label $\vec{x}$ and $\vec{q}_i$ the position of the TP and of the $i$-th bath mode, respectively.
As for the form of the interaction between particle and bath, this is a relativistic generalization of the well known Caldeira-Leggett coupling introduced in Refs. \cite{Petrosyan_2022,Zadra}.

While the factor $\gamma^{-1}$ for the kinetic energy term in the Lagrangian is standard throughout the literature, the same factor for the elastic (potential energy term) in the above Eq. \eqref{interaction} is less common. The origin of this factor, in our treatment, resides in the fact that the oscillators' spring constant $\kappa \equiv \omega^2 m$ may refer to springs connecting particles that do not necessarily travel in the same reference frame, but may belong to frames travelling at a relative speed with respect to each other. For example, this is for sure the case for the tagged particle in its dynamical connection with the bath oscillators. Since the spring constant $\kappa$, in classical mechanics, is given by a force divided by a length, using the proper length leads to a Lorentz factor $\gamma$ in the denominator. Of course, whenever the particles connected by the springs move in the same frame, then the Lorentz factor is identically equal to one, and we retrieve the standard notation without this factor. Furthermore, this factor $\gamma(\dot{x})$ in the denominator of the potential energy part of the Lagrangian Eq. \eqref{interaction} is also needed to correctly recover the proper momentum inside the memory integral for the viscous frictional force in the resulting relativistic Langevin equation \cite{Petrosyan_2022,Zadra}.

The bath oscillators $q_{i}$ are entities which mimic the degrees of freedom of the particles in a fluid, which are dynamically connected (i.e. "interacting") with the tagged particle. This is the essence of the Caldeira-Leggett model as discussed in many papers and textbooks (e.g. Weiss \cite{Weiss} or Nitzan \cite{Nitzan}). In particular, Nitzan in his monograph \cite{Nitzan} discusses in detail the mathematical mapping between the bath oscillators $q_i$ and the actual particles forming the fluid (in the non-relativistic case). 
While the $q_i$ are not exactly the particles forming the fluid, there is a mathematical mapping from the $q_i$s to the particles that constitute the fluid. It is important to recognize that the bath oscillators must be treated as relativistic oscillators to ensure that the whole fluid behaves as a relativistic fluid, since this also changes the corresponding fluctuation-dissipation theorem as discussed in \cite{Petrosyan_2022,Zadra}.

One should also note that this is a microscopic Lagrangian for classical particles, and we adopt the methods and viewpoint of microscopic theories of statistical mechanics. This allows one, among other things, to circumvent the well-known problem of field theories where the fact that the Galileo group is not a sub-group of the Lorentz group makes it difficult to retrieve the non-relativistic limit in an easy way from the relativistic case \cite{Dore}. Also, since this is a microscopic Lagrangian (and not a hydrodynamic one), it does not suffer from the problems of hydrodynamic Lagrangians discussed in Ref. \cite{Torrieri_2024} and is generally applicable to both isotropic and anisotropic systems.

One arrives at the following relativistic Langevin equation \cite{Zadra}:
\begin{equation}
		\frac{d}{dt}p^\mu=F_\mathrm{ext}^\mu+F_p^\mu-\frac{1}{m}\int_0^t K{^\mu}_{\nu}(t,s) p^\nu(t-s)ds
		\label{eq:covlang}
\end{equation}
where $K{^\mu}_{\nu}(t,s)=\mathrm{diag}(0,K'^1(t,s),K'^2(t,s),K'^3(t,s))$ is the rank-two friction tensor, $p^\mu$ is the TP's 4-momentum, $F_p^\mu$ is the stochastic force (thermal noise) due to collisions, and $F_\mathrm{ext}$ is the force due to conservative elastic interactions \cite{Petrosyan_2022,Zadra}. The viscous frictional dissipative force (third term on the r.h.s.) describes the loss of momentum due to collisions with the other gas particles.

As discussed in \cite{Zadra}, the memory function $K{^\mu}_{\nu}(t)$ cannot be reduced to a Dirac delta of time (i.e. to white noise). This is because the only way to mathematically reduce the memory function to a Dirac delta is via imposing the same value of coupling constant $g_i$ between the TP and each of the $i$-th modes of the bath. This, in turn, clearly violates the principle of locality, and is therefore incompatible with special relativity. This implies that the relativistic Langevin dynamics is intrinsically non-Markovian \cite{Petrosyan_2022}.

\subsection{From the relativistic Langevin equation to a general form for the relativistic viscosity}
As shown in Ref. \cite{zaccone2024}, Eq. \eqref{eq:covlang} can be reduced to the following relativistic Langevin equation for the (mass-rescaled) particle displacement $\mathbf{s}_{i}$ (a 3-vector):
\begin{equation}
\gamma\frac{d^2\mathbf{s}_i}{dt^2}+\int_{-\infty}^{t}\nu(t-t')
\,\gamma\frac{d\mathbf{s}_i}{dt'}dt'+\mathbf{H}_{ij}\mathbf{s}_{j}=\mathbf{\Xi}_{i,\kappa\chi}\mathbf{\bm{\eta}}_{\kappa\chi},
\label{2.4gle2}
\end{equation}
where $\gamma$ is the Lorentz factor and we take an isotropic frictional memory-function, $K^{\mu}_{\nu}(t)\equiv \nu(t)$. The particle displacements $\mathbf{s}_{i}$ are measured with respect to the original undeformed or unstrained configuration of the system.
Furthermore, $\mathbf{H}_{ij}$ represents the Hessian matrix, defined as
\begin{equation}
\mathbf{H}_{ij}=\frac{\partial U}{\partial \mathring{\mathbf{r}}_{i} \partial \mathring{\mathbf{r}}_{j}}\bigg\rvert_{\gamma \rightarrow 0} = \frac{\partial U}{\partial \mathbf{r}_{i} \partial \mathbf{r}_{j}}\bigg\rvert_{\mathbf{r}\rightarrow \mathbf{r}_{0}}\label{ring_Hessian}
\end{equation}
since $\mathring{\mathbf{r}}(\gamma)\rvert_{\gamma \rightarrow 0}=\mathbf{r}_{0}$, where $U$ denotes the total free energy of the system.

For the term on the r.h.s. of Eq. \eqref{2.4gle2} we have the following identification:
\begin{equation}
    \mathbf{\Xi}_{i,\kappa\chi}=\frac{\partial \mathbf{f}_{i}}{\partial \mathbf{\bm{\eta}}_{\kappa\chi}}
\end{equation}
and the limit $\mathbf{\bm{\eta}}_{\kappa\chi} \rightarrow 0$ is implied. Here $\eta_{\kappa\chi}$ are the components of the Cauchy-Green strain tensor defined as $\bm{\eta} =\frac{1}{2}\left(\mathbf{F}^{T}\mathbf{F}-\mathbf{1} \right)$, where $\mathbf{F}$ denotes the deformation gradient tensor \cite{landau1959fluid}. This is a second-rank tensor, not to be confused with the fluid viscosity $\eta$ (a scalar). Furthermore, $\mathbf{f}_{i}$ in the above expression denotes the force acting on the particle $i$ in its affine position, i.e. the position prescribed by the external strain tensor $\bm{\eta}$. This force has to be equilibrated for the deformation to evolve along a pathway of mechanical equilibrium. The equilibration of this force $\mathbf{f}_{i}$ gives rise to extra displacements (on top of the affine displacements dictated by $\bm{\eta}$) known as \emph{nonaffine} displacements.

From a physical point of view, $\mathbf{\Xi}_{i,\kappa\chi}$ represents the force acting on the particle in the so called affine position. The latter is the spatial position dictated by the external shear field for each value of applied shear strain, which is very different from the actual position reached by the particle due to multiple collisions with other particles, leading to nonaffine motions.

One should note that, in the above framework, we made use of the assumption of a well-defined particle number $N$, since particles are always perfectly conserved in classical special relativity. While particle number conservation is, in general, incompatible with high-energy processes described by quantum fields \cite{Srednicki}, this is still a valid assumption as long as one does not change from one inertial frame to the other. 

In linear response theory, for a viscous fluid, the stress is related to the strain via
\begin{equation}
    \sigma(t)=G'\epsilon_{xy}(t) + \frac{G''}{\omega}\dot{\epsilon}_{xy}(t) \approx \frac{G''}{\omega}\dot{\epsilon}_{xy}(t)\label{fundamental}
\end{equation}
where $\epsilon_{xy}\equiv \bm{\eta}_{xy}$ denotes the macroscopic shear strain. For a gas, the elastic modulus is basically zero, $G'=0$.
For a relativistic off-diagonal shear deformation $\mu\nu=ik$ with $i \neq k$, the second term in the above equation represents the dissipative stress. Let us recall that, for a viscous relativistic fluid, at the level of special relativity, the (viscous) stress tensor is given by  
\cite{landau1959fluid}:
\begin{equation}
\begin{split}
\sigma' \equiv \sigma_{ik}'&=c \eta \left(\frac{\partial u_{i}}{\partial x^{k}}+\frac{\partial u_{k}}{\partial x^{i}}-u_{k}u^{l}\frac{\partial u_{i}}{\partial x^{l}}-u_{i}u^{l}\frac{\partial u_{k}}{\partial x^{l}}\right)+\\
&-c (\zeta -\frac{2}{3}\eta)\frac{\partial u^{l}}{\partial x^{l}}(g_{ik}-u_iu_k),\label{relativistic_stress_0}
\end{split}
\end{equation}
with the Minkowski metric $g_{ik}=\text{diag}(+1-1-1-1)$ and $u_i$ the 4-vector displacement field \cite{landau1959fluid}.
Since we are interested in a shear deformation, in the above equation the second term vanishes, since $\frac{\partial u^{l}}{\partial x^{l}}=0$, and only the first term survives:
\begin{equation}
\sigma' \equiv \sigma_{ik}'=c \eta \left(\frac{\partial u_{i}}{\partial x^{k}}+\frac{\partial u_{k}}{\partial x^{i}}\\
-u_{k}u^{l}\frac{\partial u_{i}}{\partial x^{l}}-u_{i}u^{l}\frac{\partial u_{k}}{\partial x^{l}}\right).\label{relativistic_stress}
\end{equation}

The corresponding off-diagonal relativistic strain rate is thus:
\begin{equation}
\dot{\epsilon}_{xy} \equiv c\left(\frac{\partial u_{i}}{\partial x^{k}}+\frac{\partial u_{k}}{\partial x^{i}}-u_{k}u^{l}\frac{\partial u_{i}}{\partial x^{l}}-u_{i}u^{l}\frac{\partial u_{k}}{\partial x^{l}}\right)\label{relativistic_strain_rate}.
\end{equation}
For a gas, the total
stress reduces to the viscous stress, $\sigma \approx \sigma'$. For comparison, this is equivalent to an off-diagonal space-like component of the
stress-energy tensor, such as $\sigma = \sigma' = T^{12}$ where $T^{12}$ is the $\mu = 1, \nu = 2$
component of the stress-energy tensor $T^{\mu\nu}$ with $ \mu, \nu= 0, 1, 2, 3$. This component describes shear deformation in the spatial 12 plane of the Minkowski
space.

In a fluid (gas or liquid) that lacks shear rigidity, the Hessian matrix introduced above has to be diagonalized for each instantaneous frame \cite{Keyes_INMs,Stratt_INMs}. The diagonalization yields normal modes \cite{Kardar_2007}, which include the acoustic frequencies (positive eigenvalues of the Hessian) plus overdamped modes (negative eigenvalues) known as instantaneous normal modes \cite{Keyes_INMs,Stratt_INMs}. In all cases, the eigenvalues and eigenfrequency are related via $\lambda_p = m \omega_{p}^{2}$, where the index $p$ labels the mode \cite{Keyes_INMs,Kardar_2007}.  

Upon Fourier transforming Eq. \eqref{fundamental}, the above equations yield the following identification \cite{Landau_elasticity,landau1959fluid,lakes_2009}:
\begin{equation}
    \eta = \frac{G''}{\omega} \label{viscosity}
\end{equation}
between the viscosity $\eta$ of the system and the loss modulus $G''$. Upon performing normal mode decomposition and Fourier transformation on Eq. \eqref{2.4gle2}, the $p$-th normal mode can be written as:
\begin{equation}
-\gamma\,\omega^2\hat{\tilde{s}}_p(\omega)+i\gamma\tilde{\nu}(\omega)\omega\,\hat{\tilde{s}}_p(\omega)
+\omega_p^2\hat{\tilde{s}}_p(\omega)
=\hat{\Xi}_{p,\kappa\chi}(\omega)\tilde{\bm{\eta}}_{\kappa\chi},
\label{2.4kerneldecom}
\end{equation}
where $\tilde{\nu}(\omega)$ is the Fourier transform of the friction kernel $\nu(t-t')$ (memory function) of the generalized Langevin equation Eq. \eqref{2.4gle2}. Hence, $\tilde{\nu}(0)\equiv \tilde{\nu}(\omega=0)$ is the zero-frequency limit of the Fourier-transformed friction kernel.
Since $\hat{\Xi}_{p,\kappa\chi}=\mathbf{v}_p\cdot\mathbf{\Xi}_{\kappa\chi}$ is self-averaging \cite{Lemaitre}, one might introduce the smooth correlator function on eigenfrequency shells
\begin{equation}
\Gamma_{\mu\nu\kappa\chi}(\omega_p)=\langle\hat{\Xi}_{p,\mu\nu}\hat{\Xi}_{p,\kappa\chi}\rangle_{\omega_p\in\{\omega,\omega+d\omega\}} \label{Gamma}
\end{equation}
on frequency shells.
Following the general procedure of ~\cite{Lemaitre} to find the oscillatory stress for a dynamic deformation, the stress is obtained to first order in strain amplitude as a function of $\omega$ (note that the summation convention is active for repeated indices):
\begin{align}
\tilde{\sigma}_{\mu\nu}(\omega)&=-\frac{1}{\mathring{V}}\sum_{p}\hat{\Xi}_{p,\mu\nu}\hat{\tilde{s}}_p(\omega) \notag\\
&=-\frac{1}{\mathring{V}}\sum_{p}\frac{\hat{\Xi}_{p,\mu\nu}\hat{\Xi}_{p,\kappa\chi}}{\omega_{p}^2-\gamma\omega^2
+i\gamma\tilde{\nu}(\omega)\omega}\tilde{\bm{\eta}}_{\kappa\chi}(\omega)\notag\\
&\equiv C_{\mu\nu\kappa\chi}(\omega)\tilde{\bm{\eta}}_{\kappa\chi}(\omega).\label{generalized}
\end{align}
where, again, we neglected the elastic part of the response. Replacing the discrete sum over modes with an integral over the density of states (DOS) and specializing to off-diagonal shear deformations, $\mu\nu\kappa\chi=xyxy$, we obtain the following relativistic expression for the complex shear modulus $G^{*}$:
\begin{equation}
G^{*}(\omega)=\frac{1}{\mathring{V}}\int_0^{\omega_{D}}\frac{g(\omega_p)\Gamma_{xyxy}(\omega_p)}{m\gamma\omega^2-m\omega_p^2-i\gamma\tilde{\nu}(\omega)\omega}d\omega_p.
\label{2.4shear_modulus}
\end{equation}
The low-frequency behaviour of the correlator $\Gamma(\omega_p)$ can be evaluated analytically using the following result \cite{Scossa}:
\begin{equation}
\langle\hat{\Xi}_{p,\mu\nu}\hat{\Xi}_{p,\kappa\chi}\rangle = d\kappa
R_0^2\, \lambda_{p}\sum_{\alpha}B_{\alpha,\mu\nu\kappa\chi},\label{correlator}
\end{equation}
which gives $\langle \hat{\Xi}_{p,xy}^{2}\rangle \propto \lambda_{p}$, thus implying (from its definition above): 
$\Gamma(\omega_p) \propto \omega_{p}^{2}$.
Furthermore, $\sum_{\alpha}B_{\alpha,\mu\nu\kappa\chi}$ is a geometric coefficient that depends only on the geometry of deformation, e.g. it is equal to $1/15$ for shear \cite{Scossa}.

For shear deformation $\kappa\chi=xy$, by taking the imaginary part of Eq. \eqref{2.4shear_modulus}, the nonaffine response theory developed from first principles above thus provides the following expression for the loss modulus $G''$:
\begin{equation}
G''(\omega)=\frac{1}{\mathring{V}}\int_{0}^{\omega_{D}}\frac{g(\omega_p)\,\Gamma(\omega_p)\,\gamma\,\tilde{\nu}(\omega)\,\omega}{m^{2}(\omega_{p}^{2}-\gamma\omega^{2})^{2}+\gamma^2\tilde{\nu}(\omega)^{2}\omega^{2}}d\omega_p. \label{loss_repeated}
 \end{equation}

The zero-frequency shear viscosity $\eta = \lim_{\omega \rightarrow 0}G''/\omega$, is thus given by the following expression:
\begin{equation}
    \eta = \frac{1}{\mathring{V}}\,\gamma\,\tilde{\nu}(0)\int_{0}^{\omega_{D}}\frac{g(\omega_p)\,\Gamma(\omega_p)}{m^{2}\omega_{p}^{4}}d\omega_p. \label{viscosity_nonaffine}
\end{equation}

\subsection{Non-relativistic fluid}
In Ref. \cite{zaccone2023} using the non-relativistic Langevin theory, the following expression for non-relativistic classical systems was obtained:
\begin{equation}
    \eta_{\gamma\rightarrow 1} = \frac{1}{\mathring{V}}\,\tilde{\nu}(0)\int_{0}^{\omega_{D}}\frac{g(\omega_p)\,\Gamma(\omega_p)}{m^{2}\omega_{p}^{4}}d\omega_p. \label{viscosity_nonaffine_old}
\end{equation}
which differs from the relativistic formula Eq. \eqref{viscosity_nonaffine} only due to the extra factor $\gamma$.
Hence, we see at once that, upon taking the non-relativistic limit of Eq. \eqref{viscosity_nonaffine}, i.e. $\gamma \rightarrow 1$, the correct non-relativistic expression Eq. \eqref{viscosity_nonaffine_old} is recovered.

The presence of the Lorentz $\gamma$ factor in Eq. \eqref{viscosity_nonaffine} is ultimately due to the fact that, in the relativistic Langevin equation, the dissipative viscous force is written in terms of the proper momentum, cfr. Eqs. \eqref{2.4gle2} and \eqref{eq:covlang}.
From basic kinetic theory \cite{Born_atomic,Vincenti}, the viscosity is directly proportional to the rate of momentum loss in collisions. Within the Langevin picture \cite{Zwanzig}, the rate of momentum loss is, in turn, proportional to the momentum of the particles, $p/m$. The above derivation and result suggest that, for a gas of relativistic particles, the viscosity (as measured in the lab frame) is proportional to the \emph{proper momentum}, hence, with a relativistic enhancement by a factor $\gamma$. This is a new relativistic effect that, to our knowledge, has never been discussed before.

Since in the low-frequency sector the main excitations are (longitudinal) acoustic waves with a linear dispersion relation, we take a Debye-type approximation for the DOS spectrum, $g(\omega_{p}) = \frac{\omega_{p}^2\,V}{2\,\pi^2\, c_s^3}$, where $c_s$ is the speed of sound.

Upon recalling Eq. \eqref{correlator} and $\lambda_p = m \omega_{p}^2$, the correlator $\Gamma(\omega_p)$ is given by $\Gamma(\omega_p) = \frac{1}{5} m\kappa R_{0}^{2} \omega_{p}^{2} $ \cite{Scossa}. We thus get:
\begin{equation}
    \eta =  \frac{1}{10\pi^2}\frac{\kappa R_{0}^{2} \,\tilde{\nu}(0)}{c_s^3} \frac{\omega_{D}}{m}.\label{simplified2}
\end{equation}

We shall now proceed to showing how the above relativistic Eq. \eqref{viscosity_nonaffine} recovers the well known temperature scaling of the viscosity of classical gases in the non-relativistic limit, $\gamma =1$.

In a gas of classical hard spheres, as is known, the excluded volume is responsible for an effective weak attraction between nearby particles (a manifestation of the so-called depletion attraction \cite{Hansen}). This becomes most evident by considering the radial distribution function $g(r)$. For the hard-sphere (HS) fluid (which is always a good approximation at high temperature also in presence of interactions), the $g(r)$ is well described, especially at low density, by the Percus-Yevick theory \cite{Hansen,Dhont,AZET}. In the low-density HS fluid of interest here, the $g(r)$ features a small peak near contact, which rapidly decreases towards $g(r)=1$ (cfr. p.4 in \cite{Barrat_Hansen_2003}, Fig. 7 in \cite{Bolmatov} or Fig. 4 in \cite{Widom}). As was found by Verlet and Weis \cite{Verlet} (see also \cite{Henderson}), the $g(r)$ near particle contact is well parameterized by the following expression valid for $r>d$, with $d$ the particle diameter \cite{Henderson}:
\begin{equation}
    g(r/d)= g(1^+)+ \frac{C}{r}\exp[-m(r-d)] \label{verlet}
\end{equation}
where $g(1^+)$, $C$, and $m$ are parameters which depend only on the density and we have neglected the oscillating factor $\sim\cos[m(r-d)]$, representing Friedel oscillations, since these oscillations are absent in the HS gas and show up only at higher density in the liquid. Furthermore, $g(r)=0$ for $r<d$ (impenetrability condition of hard spheres).
According to Eq. \eqref{verlet}, the $g(r)$ starts off at $r/d=1$ at some finite value $g(1^+)$ and then decays to $1$. The corresponding potential of mean force (pmf) describes the effective interaction between two particles mediated by all the other particles in the system \cite{Chandler}. The pmf is given by \cite{Chandler}:
\begin{equation}
    V_{\mathrm{pmf}}(r)=-k_B T \ln g(r).
\end{equation}
Since there is a peak near contact in the $g(r)$, the pmf $V_{\mathrm{pmf}}(r)$ has an attractive well (i.e. the depletion attraction mentioned above) corresponding to the peak in $g(r)$. The effective spring constant corresponding to this interaction energy minimum is therefore: $\kappa \approx k_B T/d^2$.

Furthermore, for the maximum vibrational frequency of sound in the system, $\omega_D$, this is proportional to the vibrational temperature $\omega_D =k_B T/h$, where $T$ is the temperature of the gas. Finally, the speed of sound in a gas is given by: $c_s \propto \sqrt{k_BT/m}$ \cite{Vincenti}.
Upon substituting $\kappa \approx k_B T/d^2$, together with $\omega_D \sim T$ and $c_s \propto \sqrt{k_BT/m}$ in Eq.\eqref{simplified2}, we get:
\begin{equation}
\eta \propto \frac{\sqrt{m k_B T}}{d^2}.\label{classic}
\end{equation}
This is quite a surprising and important outcome. In spite of starting from a very different framework (the relativistic Langevin equation and linear response theory), we have recovered the same dependence of viscosity on temperature, mass, Boltzmann's constant, and particle size $d$, which is predicted by the kinetic theory of gases \cite{Vincenti,Born_atomic,Hildebrand}: $\eta \propto \frac{\sqrt{m k_B T}}{d^2}$, barring constant prefactors. Furthermore, this is the first time that the viscosity of classical non-relativistic gases is derived starting from a relativistic formula, and, in fact, from a relativistic Lagrangian.

\subsection{Relativistic and ultrarelativistic limit}
We shall now consider the opposite limit of a dense, strongly correlated high-energy fluid of relativistic or ultrarelativistic particles (as such, a model for the quark gluon plasma). Since the system, also in this case, supports acoustic waves with a linear dispersion relation between frequency $\omega_p$ and momentum $p$, we can keep working with a DOS quadratic in frequency (momentum). This is also what one obtains from relativistic statistical mechanics in the ultrarelativistic limit \cite{Pathria}: $g(p) =\frac{4\pi V}{h^3 c^3} (cp)^2$, barring prefactors, with $p$ the momentum.

Hence, working with frequency, we take a parabolic approximation for the DOS, $g(\omega_{p}) = \frac{V\,\omega_{p}^2}{2\,\pi^2\, c_s^3}$, where $c_s$ is the speed of sound.
Next we use $\lambda_p = m \omega_{p}^2$, and the correlator $\Gamma(\omega_p) = \frac{1}{5} m\kappa R_{0}^{2} \omega_{p}^{2} $ \cite{Scossa} in Eqs. \eqref{viscosity_nonaffine} and \eqref{correlator}, and we thus get the relativistic version of Eq. \eqref{simplified2}:
\begin{equation}
    \eta =  \frac{1}{10\pi^2c_s^3}\gamma\kappa R_{0}^{2} \,\tilde{\nu}(0) \frac{\omega_{D}}{m}.\label{simplified3}
\end{equation}
Again, the effective spring constant $\kappa$ has dimensions of force per unit length, and in the high-T QGP-type fluid, dominated by collisional physics, it is set by the relevant energy scale divided by the interparticle separation squared, $\kappa \sim \frac{k_B T}{R_{0}^2}$. Note that in this case of a high-density fluid, $R_0 \sim d$.
One should also note that now the Lorentz factor is not unity, but is proportional to temperature, $\gamma \sim T$, as appropriate for an ultrarelativistic fluid (since $\gamma m c^2 \approx k_B T$). Furthermore, in the regime $T>2 T_c$, where $T_c$ is the Hagedorn temperature, we can assume the speed of sound $c_s$ of the QGP to be approximately $T$-independent and approximately equal to $c/3$ \cite{BEGUN_2011}. Therefore, we obtain the following estimate for the temperature dependence of the viscosity of an ultrarelativistic dense fluid:
\begin{equation}
    \eta =  \frac{1}{10\pi^2}\frac{\tilde{\nu}(0)}{m^2}\frac{(k_BT)^3}{h c^5} ,\label{simplified4}
\end{equation}
and in the case of QGP one would need to put in an extra factor 12 to account for the 3 colors, 2 spin states and particle-antiparticle degeneracy. 
This is a new analytical formula for the viscosity of an ultrarelativistic gas in terms of fundamental constants such as the Boltzmann's constant, the speed of light, and the Planck's constant, and the temperature and mass of the particles.
As a sanity check, the dependence of the viscosity on the cube of temperature is backed by evidence on high-energy fluids and various theories and estimates for the QGP viscosity \cite{Baym,Redlich,Shuryak_RMP,kapusta_gale_2023}.

The regime of applicability of the above formula should be discussed. The result Eq. \eqref{simplified4} applies to classical (non-degenerate) hard particles at relativistic speeds (with a corresponding dispersion relation, cfr. \cite{Pathria}) in a regime where (i) particle-number changing processes are neglected (as they are in classical treatments of ultrarelativistic fluids, cfr. \cite{Pathria} and \cite{Cerci}), (ii) temperature is very high, $\zeta = m c^2/(k_B T) \ll 1$ by definition of ultrarelativisitc limit, such that quantum effects are negligible (classical systems). Furthermore, (iii) the gas is dense such that multi-particle collisions are significant. The latter point explains the difference with the formula obtained from the relativistic Boltzmann equation (cfr. Eq. 5.100 on page 125 of Cercignani \& Kremer's book \cite{Cerci}), which is only linear in $T$, possibly because it is derived under a Chapman-Enskog scheme. Hence, it can be said that the formula Eq. \eqref{simplified4} is an improvement with respect to the corresponding formula obtained from the relativistic Boltzmann equation, as it correctly predicts the $T^3$ behaviour of hot dense matter (in agreement with other approaches, e.g. \cite{Denicol}).
It remains questionable that the regime outlined above is possible in relativistic systems with no conservation of charges and particles. This is 
because radiative processes in the intermediate ($P \sim T$) to soft ($P \ll T$)
channels will dominate over multi-particle collision processes. These issues were tackled e.g. in the work of  Arnold, Moore and Yaffe \cite{Arnold} at the level of a field theory, which, through resummation, leads to a Boltzmann equation.

\section{Comparison with the relativistic Boltzmann equation theory}
The other main approach, to our knowledge, which can produce analytical formulae for the viscosity across the whole energy spectrum, i.e. from the non-relativistic classical gas to the ultrarelativistic gas of hard particles, is the relativistic Boltzmann equation theory \cite{Cerci}.
Although, in general, the relativistic Boltzmann equation has to be solved numerically, analytical limits exist. These are obtained by using kernels in the form of modified Bessel functions in the parameter $\zeta = m c^2/(k_B T) $.
As shown in \cite{Cerci}, in the non-relativistic limit for hard particles, the relativistic Boltzmann equation yields the following analytical formula
\begin{equation}
    \eta = \frac{5}{16 d^2}\left(\frac{m k_B T}{\pi}\right)^{1/2}
\end{equation}
where, in Cercignani and Kremer's notation, $d$ is the hard-sphere particle diameter. The above equation clearly recovers the well-known result of the kinetic theory of gases \cite{Vincenti} and also agrees with the analytic limit we derived above for the non-relativistic dilute gas with our new method, Eq. \eqref{classic}.\\
In the opposite limit of ultrarelativistic, $\zeta \ll 1$, hard particles, the relativistic Boltzmann equation yields the following analytical formula:
\begin{equation}
\eta = \frac{3}{10 \pi}\frac{k_B T}{c \sigma}\label{cerci}
\end{equation}
to first order in a power series expansion in $\zeta$. In the above expression, $\sigma$ represents the differential cross-section of the gas of hard-sphere particles.
We notice that the predicted temperature-dependence is only linear in $T$. This differs from our result Eq. \eqref{simplified4}, which predicts a cubic dependence on $T$ in agreement with theoretical approaches for QGP and weakly-interacting gases \cite{Denicol}. A reason for this difference can be ascribed to the fact that Eq. \eqref{cerci} is derived from the relativistic Boltzmann equation by making use of the Chapman-Enskog approximation, while Eq. \eqref{simplified4} has been derived above for a dense gas where the nearest-neighbor distance is of the same order of the particle diameter.

\section{Conclusions}
In conclusion, we have presented a microscopic derivation of the viscosity of relativistic fluids, which is able to correctly recover the known result from the kinetic theory of gases in the non-relativistic limit in a fully analytical way. In the limit of ultrarelativistic fluids, the theory produces a new formula, with a cubic temperature dependence, which provides a new fundamental law bringing together the Boltzmann's and Planck's constants, the speed of light and the mass and temperature of the particles. This temperature dependence is different from the linear one predicted by the relativistic Boltzmann equation theory (with the Chapman-Enskog scheme) in its analytic limit for ultrarelativistic hard-sphere gases, because it accounts for multi-particle collisions and is thus more suitable to describe dense gases.
In the future, this theory can be extended in several directions, e.g. to obtain similar predictions for the bulk viscosity \cite{Gavassino_2021} with possible applications to neutron stars \cite{PhysRevD.107.103031}, and as an alternative to Kubo formulae for hot particle gases \cite{Kaiser}.
Since the viscosity is derived from a Langevin equation, it should also be possible to evaluate the associated entropy production \cite{Landi} and compare with the entropy density for the relativistic fluids from field theories and other estimates \cite{Kapusta_PRL}.
Finally, the present approach could be applied to plasmas, where quasilinear theory (QLT) \cite{Besse_2011} has represented a successful paradigm and its recent developments allow for calculations of viscosity for both relativistic and non-relativistic plasmas \cite{Dodin}. Since the current approach is fundamentally different from QLT (in the same way as the generalized Langevin equation picture differs fundamentally from the Fokker-Planck picture), it will be very interesting to explore the application of the present theory to realistic plasma systems.

\subsection{Acknowledgments} 
I am indebted to Dr. Lorenzo Gavassino (Vanderbilt University) for input and many useful discussions. I also would like to thank Prof. Matteo Baggioli (Shanghai Jiao Tong University) for a preliminary reading of the manuscript.
I gratefully acknowledge funding from the European Union through Horizon Europe ERC Grant number: 101043968 ``Multimech'', from US Army Research Office through contract nr. W911NF-22-2-0256, and from the Nieders{\"a}chsische Akademie der Wissenschaften zu G{\"o}ttingen in the frame of the Gauss Professorship program. 

\bibliographystyle{apsrev4-1}

\bibliography{refs}

\end{document}